\documentclass[aip, jcp, amsfonts, amssymb, amsmath, reprint, showkeys, nofootinbib, twoside]{revtex4-1}
\usepackage{xcolor}
\usepackage{graphicx}
\usepackage{wrapfig}
\usepackage{comment}
\usepackage{mathtools} 
\newcommand{\wav}{\,\ensuremath{\mathrm{cm}^{-1}}}
\usepackage{mhchem}

\usepackage{bm}
\newcommand{\vect}{\boldsymbol}
\newcommand{\polymer}[1]{\mathbf{#1}}

\newcommand{\e}{\mathrm{e}}
\newcommand{\x}{\tilde{x}}

\DeclarePairedDelimiterX\braket[2]{\langle}{\rangle}{#1 \delimsize\vert #2}

\usepackage[pdftex, pdftitle={Article}, pdfauthor={Author}]{hyperref} 
\renewcommand{\d}{\mathrm{d}}
\begin{document}

\title{High-Accuracy Molecular Simulations with
  Machine-Learning Potentials and Semiclassical Approximations to Quantum Dynamics}

\author{Valerii Andreichev}
\altaffiliation{These authors contributed equally}
\affiliation{Department of Chemistry, University of Basel, CH-4056 Basel, Switzerland}
\author{Jindra Du\v{s}ek}
\altaffiliation{These authors contributed equally}
\affiliation{\mbox{Department of Chemistry and Applied Biosciences, ETH Zurich, CH-8093 Zurich, Switzerland}}
\author{Markus Meuwly}
\email[Correspondence email address: ]{m.meuwly@unibas.ch}
\affiliation{Department of Chemistry, University of Basel, CH-4056 Basel, Switzerland}
\author{Jeremy O. Richardson}
\email[Correspondence email address: ]{jeremy.richardson@phys.chem.ethz.ch}
\affiliation{\mbox{Department of Chemistry and Applied Biosciences, ETH Zurich, CH-8093 Zurich, Switzerland}}

\begin{abstract}

Accurate simulations of molecules require high-level electronic-structure theory in combination with rigorous methods for approximating the quantum dynamics.
Machine-learning approaches can significantly reduce the computational expense of this workflow without any loss of accuracy.
We discuss various methods for constructing potential energy surfaces including  transfer learning, which requires a minimal number of expensive training points.
In this way, we can study chemical reactions at a high level but a low cost.
In particular, as the potentials are smooth and differentiable, they enable the use of more advanced semiclassical approximations to quantum dynamics, such as perturbatively corrected instanton theory, which can capture both tunnelling and anharmonicity.
\end{abstract}


\maketitle

\section{Introduction}
Quantitative molecular simulations require both accurate potential energy surfaces (PESs) and 
accurate quantum-dynamics methods. Unfortunately, highly accurate {\it ab initio}
electronic-structure calculations are computationally extremely
expensive.
For example, coupled cluster with singles, doubles and
perturbative triples, CCSD(T), which is considered the ``gold
standard'' for single-reference problems, scales as
$\mathcal{O}(N^7)$, where $N$ denotes the number of basis
functions. This means that doubling the size of the molecule requires 128 times more computational time. Simulating quantum dynamics for nuclear motion can be even harder as it scales exponentially
with the number of degrees of freedom. In this overview we discuss recent progress on a two-pronged attack of this central challenge in chemistry, culminating in the combined use of highly accurate machine-learned PESs with advanced semiclassical approximations to quantum dynamics.\\

\noindent
Constructing a PES using machine learning (ML) is a multi-step process, ranging from the formulation and definition of the problem of interest to validating the fitted model and performing further refinements.\cite{MM.rev:2021,unke2021machine,MM.rev:2023} The aim is to minimize the number of reference points required to achieve the quality level needed for the problem to be solved. One method to achieve this is ``transfer learning'',\cite{pan:2010} which elevates a model thoroughly trained at a lower level of theory to a higher level by supplying only a small number of high-level training data. This is particularly relevant for neural networks (NN) which would otherwise require considerable amounts of training samples (thousands to tens of thousands).\\

\noindent
After the machine-learning process is completed, we can run dynamical
simulations at considerably reduced computational expense even though
it is necessary to evaluate the PES for a large number of geometry
($\sim 10^6$ or more), depending on the type and purpose of the
simulation. However, despite this saving, the intrinsic cost of full
quantum dynamics can still be insurmountable. Rigorous approximations
to quantum dynamics are therefore required that scale as classical
algorithms.  For certain problems in reactive scattering,
quasi-classical trajectory simulations are a good compromise between
accuracy and efficiency.  However, they cannot capture quantum effects
such as tunnelling, which play an important role in many other
reactions of interest.  We therefore introduce the semiclassical
``instanton'' approximation for simulating quantum tunnelling in
molecules.\cite{Miller1975semiclassical,Chimia,Perspective} Recently,
we have developed a systematic improvement to this approximation by
including perturbative corrections to capture the anharmonicity of
molecular vibrations.\cite{AnharmInst} The resulting
perturbatively-corrected instanton method combined with an accurate
machine-learned PES is a powerful approach for predicting tunnelling
effects in polyatomic molecules and, as we shall show, leads to
excellent agreement with experiment.\\

\section{Machine Learned Potential Energy Surfaces for Physical Simulations}
The two machine learning-based techniques discussed in the present work are neural network (NN) and kernel representations of potential energy surfaces.\cite{MM.rev:2023,MM.rkhs:2017} These two approaches are complementary to one another in that they are often used for specific systems and applications as indicated below. However, the two techniques can also be combined such as in KerNN which, however, will not be discussed here.\cite{MM.kernn:2025}\\

\noindent
{\it Neural Networks:} One successful approach for training NN-based PESs uses graph neural
networks (GNNs)\cite{scarselli2009gnn}, particularly message-passing
neural networks (MPNNs)\cite{gilmer2017mpnn}, whereby a molecule is
represented as a graph with atoms as nodes and bonds as edges. A
suitable atomic representation (embedding) is then directly trained from fitting the weights and biases of the NN to
data.\cite{MM.physnet:2019,unke2021machine} In PhysNet, each atom is represented as a node with a learnable embedding initialized according to its atomic number $Z$. Atoms within a predefined cutoff radius (typically 6 \AA\/) are connected by edges that encode interatomic distances through a radial basis expansion, without relying on predefined chemical bonds or bond orders. Hence, such a NN-architecture is by definition suitable for describing chemical reactions akin to {\it ab initio} MD simulations. Through iterative message-passing, atomic representations are learned which describe each atom's local geometric environments. Finally, the total energy of the molecule(s) considered is obtained as a sum of atomic energy contributions, possibly augmented by learned electrostatic interactions.\cite{MM.physnet:2019} \\

\noindent
Learnable descriptors are the result of an optimisation process which minimizes a given loss function. For PhysNet this is a weighted squared loss including energies, forces, and molecular dipole moments. To fully specify the problem, additional (user-specified) hyperparameters are required. These include, for example, the relative weights in the loss function, the number of layers the NN consists of, the length of the atomic descriptors, the type and number of basis functions used to encode the spatial
coordinates of the system, or the radial cutoff which defines the near- or far-sightedness of a trained model.\cite{MM.rev:2023,MM.rev:2025} All these quantities can be varied and model performance may depend on the choices made.\cite{MM.physnet:2019}\\

\noindent
The cornerstone for constructing a ML-PES is the data used to train
it. An open challenge in the field concerns the number of data points
and the geometrical structures chosen to cover configurational
space for the system of interest to obtain a robust ML-PES for a single chemical species. More broadly speaking: it is of great and future interest to make ML-PESs for individual chemical species also suitable for related, but different chemical compounds. This is also referred to as ``chemical transferability'' which is a great challenge due to the vast size of chemical space.\cite{reymond:2012,avl:2013} However, obtaining complete coverage of chemical {\it and}
conformational space for the construction of ML-PES is an impossible
task; therefore, biases in databases are
inevitable.\cite{wang2025design}\\

\noindent
Although for small molecules, one can train directly on CCSD(T) data,\cite{Wang2008malonaldehydePES}
obtaining a sufficiently large number of data points for
molecules of the size of tropolone (9 heavy atoms) at the
CCSD(T)/aug-cc-pVTZ level of theory is very challenging or even
impossible without further approximations.\cite{nandi2023ring}
Therefore, a solution based on using
transfer learning (TL)\cite{pan:2010} to improve a base ML-PES constructed at a modest
level of theory (MP2) to the required level of theory (CCSD(T)) using
a minimal number of high-level calculations selected using an
algorithm based on farthest-point sampling\cite{garcke:2023} was
employed.\cite{MM.tl:2022,MM.tl:2025} Generating a suitable
training dataset for an ML-PES is usually an iterative process and requires careful selection of the high-level training points which can be partly automated.\cite{MM.tl:2022} \\

\noindent
Other important aspects to consider when choosing a neural-network
architecture include floating-point precision and the programming
language. Most modern ML algorithms operate with single-precision
(32-bit) arithmetic; however, in MD simulations this can be
problematic. Recent studies\cite{MM.acc:2024} have shown that models
trained in single precision produce unreliable derivatives, yielding
rougher ML-PES surfaces compared to those obtained with double
precision (64-bit).\\

\noindent
A concrete application of an NN-based representation using PhysNet and employing transfer learning is discussed further below.\\

\noindent
{\it Reproducing Kernels:} For sufficiently small systems, kernel-based methods have emerged as
a powerful alternative to NNs.\cite{MM.h2co:2020,MM.rkhs:2020,MM.co2:2021} As a recent example for the
performance of a reproducing kernel Hilbert space (RKHS) approach, the
reaction dynamics of the O($^3$P) + O$_2$(X$^3\Sigma_g^{-} )$
$\leftrightarrow$ O($^3$P) + O$_2(^3\Sigma_g^{-} )$ system was
constructed at the MRCI+Q level of theory together with the
aug-cc-pVTZ basis set. The MRCI+Q calculations were based on
multistate CASSCF(12,9) calculations and 8 states were included in the
stat-averaged calculations. The grid for the single-channel PESs
included 2269 geometrically feasible ground-state geometries and the
three possible channels (O$_{\rm A}$O$_{\rm B}$+O$_{\rm C}$, O$_{\rm
  A}$O$_{\rm C}$+O$_{\rm B}$, and O$_{\rm B}$O$_{\rm C}$+O$_{\rm A}$)
were mixed using an exponential switching function depending on the
internuclear separations. For the crossing region, a separate data set
was generated. The single-channel PES featured a representation error
of ${\rm RMSD} < 10^{-5}$ eV (0.0002 kcal/mol) and $r^2 = 1.0$ across
9 eV whereas the mixed PES, describing all 3 asymptotic channels,
yielded RMSD = 0.047 eV (1.1 kcal/mol) and $r^2 = 0.9981$ for
``on-grid'', and RMSD = 0.13 eV $(\sim 2.9$ kcal/mol) and $r^2=0.9951$
for ``off-grid'' points.\cite{MM.o3:2025}\\

\noindent
Quasi classical trajectory (QCT) simulations using the RKHS-PES and a
PES represented as permutationally invariant polynomials (PIPs) based
on extended multi-state (XMS) complete active space second-order
perturbation theory (XMS-CASPT2) with a minimally augmented
correlation-consistent polarized valence triple-zeta (maug-vtz) basis
set based on reference states from SA-CASSCF(12,9)
calculations\cite{varga:2017} were compared with experiments. Both
sets of extensive simulations yielded very favourable agreement with
measurements for the oxygen atom-exchange reaction.\cite{wiegell:1997}
Notably, both PESs used in the present work---RKHS and PIP---feature
``reefs'' in the entrance channel. Nevertheless, the QCT-simulations
using both PESs reproduce the experimentally observed negative
$T$-dependence of the thermal rates for the exchange reaction. Hence,
the notion that the ``reef'' in earlier PESs is responsible for the
positive $T$-dependence of $k(T)$,\cite{fleurat:2003,Li:2014} which is
inconsistent with experimental observations. For the atomization
reaction O($^3$P) + O$_2$(X$^3\Sigma_g^{-} )$ $\rightarrow$ 3O($^3$P)
the $T-$dependence of the rate is correctly
described,\cite{Byron:1959,Shatalov:1973} but the magnitude depends on
the electronic degeneracy $g_e$ factor used. Assuming that the 27
possible oxygen states connect to the single ozone ground state O$_3$($^1$A) yields $g_e = 1/27$ whereas adopting $g_e = 16/3$ borrowed from
the O$_2$+Ar reaction\cite{nikitin:1974} yields rates that are two
orders of magnitude larger and in almost quantitative agreement with
measurements.\\

\section{Perturbatively Corrected Instanton Theory}
Many non-rigid molecular entities such as ammonia,\cite{Dennison1932NH3} malonaldehyde\cite{Baughcum1984malonaldehyde} or water clusters\cite{Liu1996clusters} possess two (or more)
symmetry-equivalent minimum-energy configurations. To a first approximation, this results
in degeneracy of energy levels. Quantum mechanics however, allows the system to tunnel 
between the individual configurations, which
breaks the ground-state degeneracy, giving rise to two distinct energy levels.\cite{Hund1927tunnel}
The small difference in energy levels $\Delta=E_--E_+$ can be observed spectroscopically and is called a
\emph{tunnelling splitting}. 
These splittings are known to be highly sensitive to the details of the PES\cite{StoneBook} and the validity of the approximations used to simulate the quantum dynamics.
For this reason, they make an excellent stress test for our methodology.

The molecular systems we are interested in have many degrees of freedom, such that 
solving the Schrödinger equation and obtaining $\Delta$ from the energy levels directly
would be computationally impractical. We thus seek alternative approaches.
\emph{Ring-polymer instanton theory}\cite{tunnel,Chimia,InstReview} is a method that rigorously approximates the quantum-mechanical result
with a classical computational cost and thus provides an attractive theoretical framework.

\begin{figure}[h]
\centering
\includegraphics{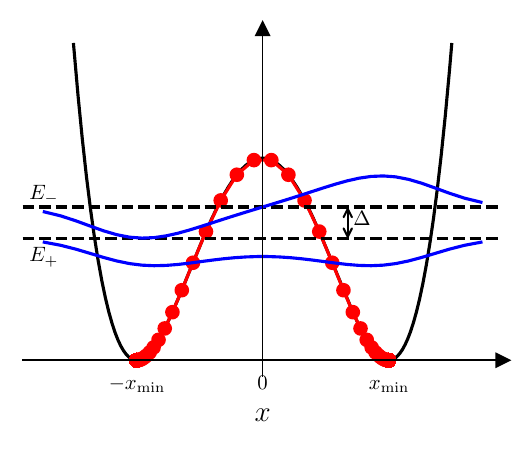}
\caption{The double-well potential $V(x) = \left( x^2/x_{\mathrm{min}}^2+1\right)^2$ is the simplest model for a system with
degenerate minima which result in a tunnelling splitting $\Delta$.
The lowest two energy levels $E_-$ and $E_+$ and wavefunctions $\psi_-(x)$ and $\psi_+(x)$ are shown.
Additionally, an instanton trajectory with $N=1024$ beads is depicted.
}
\label{fig:DoubleWell}
\end{figure}

Although in practice, we simulate molecules in full dimensionality,
in this perspective article, we will introduce our methods in the context of a simple model---the 1D double-well
depicted in Fig.~\ref{fig:DoubleWell}. 

In instanton theory, we start from an exact 
expression for the tunnelling splitting in terms of partition functions in the low-temperature limit:\cite{Benderskii}
\begin{align}
\frac{Z_\mathrm{double}}{2Z_\mathrm{single}} = \frac{\mathrm{e}^{-\beta(E_0-\Delta/2)} + \mathrm{e}^{-\beta(E_0+\Delta/2)}}{2\,\mathrm{e}^{-\beta E_0}} = \cosh\frac{\beta\Delta}{2}
\,.
\end{align}
%
Here $\beta=1/k_\mathrm{B}T$ is the inverse temperature familiar from thermodynamics.
The partition functions $Z_\mathrm{double}$ and $Z_\mathrm{single}$ correspond to a system with or without the inclusion of tunnelling.
The former is the partition function for the double-well problem,
whereas the latter is the partition function of a single well.
These expressions are evaluated in the low-temperature limit such that only the lowest vibrational states contribute.

We then express the partition functions using the \emph{discretized path-integral} formalism\cite{Feynman,Kleinert,InstReview}
\begin{align}
\label{eq:PI}
Z
&=
\lim_{N\to\infty}
\left( \frac{m}{2\pi \delta \tau \hbar} \right)^{N/2}
\int_{-\infty}^{\infty} \mathrm{e}^{-S_N(\polymer{x}, \tau)/\hbar} \, \mathrm{d} \polymer{x}
\,,
\end{align}
In this formalism, the partition function is calculated as an integral over \emph{trajectories} in imaginary time $\tau=\beta\hbar$.
The trajectories are represented by a vector $\bm{x} = ({x}_1, \dots, {x}_N)$ of $N$ \emph{beads} (and $N$ imaginary-time intervals $\delta\tau$).
Each bead ${x}_n$ is an instantaneous configuration of the molecule 
at the time $n\delta\tau$.
Lastly, the action is given by
\begin{align}
\label{eq:action}
S_N(\polymer{x}, \tau) &=
\sum_{n=1}^N 
\left[
\frac{m (x_n - x_{n-1})^2}{2\delta\tau}
+ V(x_n) \delta\tau 
\right]
\,,
\end{align}
where we employ cyclic boundary conditions ${x}_0\equiv {x}_N$.

At first sight, the multi-dimensional integral in Eq.~\ref{eq:PI} seems immensely difficult to evaluate, since the integration range spans over every possible closed path that the atoms could follow (classically allowed or not). 
However, it is clear that not all paths will contribute equally.
The most important path is the one which minimizes the action.
Rigorous asymptotic analysis\cite{BenderBook} allows the integral to be evaluated approximately based on knowledge of this single path.
This approach introduces a relative error of $\mathcal{O}(\hbar)$ and is thus known as a semiclassical approximation.
However, if we are prepared to put in extra computational effort, it is possible to reduce the error to higher orders of $\hbar$.  In this work, we will also present the perturbative corrections necessary to reduce the error to $\mathcal{O}(\hbar^2)$.

To elucidate the principle behind this semiclassical approximation, we shall consider the following simple example of a one-dimensional integral
\begin{align}
\label{eq:Ihbar}
I(\hbar) = \int_{-\infty}^\infty \e^{-f(x)/\hbar} \, \d x
\,,
\end{align}
where the function $f(x)$ has a single global minimum $\x$. We observe that the maximum of the integrand occurs at $\x$. Furthermore, the smaller $\hbar$ is in comparison to $f(x)$, the sharper the integrand becomes. Thus, as $\hbar\to0$, only the neighbourhood of $\x$ contributes to $I(\hbar)$ and it is a good approximation to express $f(x)$ in terms of its Taylor series around the minimum $\x$:
\begin{align}
\notag
f(x) &= f(\x) + \frac{1}{2!}f^{(2)}(\x)(\Delta x)^2 \\
\label{eq:fTaylor}
&+ \frac{1}{3!}f^{(3)}(\x)(\Delta x)^3 +\frac{1}{4!}f^{(4)}(\x)(\Delta x)^4 + O(\Delta x^5) \,,
\end{align}
{with $\Delta x = x-\x$.}
Note that the first derivative does not appear as $f^{(1)}(\tilde{x})=0$ at the minimum.
The second derivative $f^{(2)}(\tilde{x})$ encodes the frequencies and the higher-order derivatives encode anharmonic corrections.

After inserting Eq.~\eqref{eq:fTaylor} into Eq.~\eqref{eq:Ihbar} and using $\exp(-y)\sim1 - y + y^2/2!-\cdots$, we can further 
simplify the original integral:
\begin{align}
\notag
&I(\hbar) \overset{\hbar\to0}{\sim} \e^{-f(\x)/\hbar} \int_{-\infty}^\infty   \exp \left[- \frac{1}{2!}f^{(2)}(\x)(\Delta x)^2 /\hbar\right]  \times
\\
\label{eq:Gaussian}
&\phantom{\sim}
\left[
1
-
\frac{1}{4!\hbar}
f^{(4)}(\x) (\Delta x)^4
+
\frac{1}{2!}
\left(
\frac{1}{3!\hbar}
f^{(3)}(\x) (\Delta x)^3
\right)^2
\right]
\d x
\,,
\end{align}
where we have not explicitly written the $(\Delta x)^3$ term, as it integrates to zero, nor the $(\Delta x)^8$ term as it contributes to the second-order $\mathcal{O}(\hbar^2)$ correction.

In this way, we have approximated the intractable integral in Eq.~\eqref{eq:Ihbar} into  three standard \emph{Gaussian integrals}, which can be evaluated in closed form using well-known formulas. In the end, we obtain the simple approximation
\begin{subequations}
\label{eq:Ihbar_res}
\begin{align}
I(\hbar) &\overset{\hbar\to0}{\sim} \e^{-f(\x)/\hbar}  \sqrt{\frac{2\pi\hbar}{f^{(2)}(\x)}} \left[ 1 + \hbar \left(\Gamma_A + \Gamma_B \right)  \right]
\\
\Gamma_A &= - \frac{3}{4!} \frac{f^{(4)}(\x)}{f^{(2)}(\x)^2}
\\
\Gamma_B &= \frac{15}{2(3!)^2} \frac{f^{(3)}(\x)^2}{f^{(2)}(\x)^3}
\,.
\end{align}
\end{subequations}

As we see in Eq.~\eqref{eq:Ihbar_res}, the whole integral in Eq.~\eqref{eq:Ihbar} can be approximated using only information from the stationary point $\x$.
Apart from an extra complication due to a zero-frequency mode and a projection onto the ground-rotational state, which are dealt with in our original publication,\cite{AnharmInst}
this approach can be directly generalized for the path integral of Eq.~\eqref{eq:PI} to give a closed-form expression in terms of the stationary path $\tilde{\vect{x}}$.
This tunnelling path is called the \emph{instanton}\cite{Miller1975semiclassical} trajectory and we depict it in Fig.~\ref{fig:DoubleWell}.

While the leading-order result (ignoring the $\Gamma$ terms) requires only second derivatives of $f(x)$, the result can be improved by a perturbative correction $\hbar(\Gamma_A+\Gamma_B)$ if one has access to third and fourth derivatives of $f(x)$.
Analogously, standard instanton theory only requires hessians of the PES along the instanton trajectory and the perturbative correction captures anharmonic contributions via the third and fourth derivatives of the PES\@.
Testing this new theory on a number of systems, including one-dimensional double-well models as well as malonaldehyde,
has shown that the perturbative correction significantly improves the accuracy of the result, in many cases, by reducing the error from 20\% to 2\%.\cite{AnharmInst}

\section{Combining Instanton Theory with Machine-Learning Potentials}
The combination of ML-PESs with instanton theory represents a
transformative approach to solving the ``accuracy vs cost'' dilemma in
quantum dynamics simulations.
For example, the calculation of tunnelling splittings
using ring-polymer instanton (RPI) theory requires thousands of energy
and force evaluations due to the large number of beads required to optimize the tunnelling path. 
In addition, hundreds of Hessians are required (one for each bead) after the optimization is completed.
For the perturbative corrections, third- and fourth-order derivatives of the potential are additionally required for each bead.
Although it is possible to optimize instantons on-the-fly using methods like DFT,\cite{Hgraphene,TEMPO,methanol,nitrene,carbenes,ChimiaCarbenes} MP2\cite{chiral} or CASSCF,\cite{oxygen}
these calculations are expensive,
and performing these evaluations directly at the
CCSD(T) level of theory is computationally infeasible except for the smallest molecules.\cite{HCH4} 
For this reason, it is highly advantageous to use ML-PESs to speed-up calculations and reduce computational cost
without losing precision.

The first applications of ring-polymer instanton theory employed global ML-PESs constructed for general use, such as for water clusters.\cite{Wang2009water,Babin2014MBpol,water,hexamerprism}
However, constructing a spectroscopically accurate global PES for larger molecules remains a big challenge in general.
Combined applications of instanton theory with ML-PESs is saved by the fact that the instanton path only probes a small local region of the PES\@. 
We therefore proposed constructing a ML-PES targeting only the local region around the instanton.\cite{GPR}
This approach was based on Gaussian Process Regression (GPR) \cite{GPRbook} and only required a small number of electronic-structure calculations (energies, gradients and Hessians) computed along the instanton path, with new points added iteratively as the instanton optimization is refined.
For the case of \ce{H + CH4}, the instanton calculation converged to give the same rate (within 1\%) as the original on-the-fly calculation, despite the fact that it required only about 6 Hessians rather than 64.
This brings the cost of an instanton calculation into the same order of magnitude as a classical transition-state theory calculation, which would require at least two Hessians, one at the reactant minimum and at the saddle point (and maybe more to aid the transition-state optimization).
This GPR-instanton approach enabled the calculation of rates for reactions with larger molecules such as \ce{H + C2H6} and \ce{H + C3H8} at the level of CCSD(T), \cite{Muonium}
as well as reactions of molecules on surfaces in full dimensionality.\cite{newGPR}

Although the GPR-instanton approach works well for lowest-order instanton calculations, in order to employ the perturbative corrections, third- and fourth-order derivatives are additionally required. For computational efficiency, one would not want to compute such a high-order tensor using numerical derivatives at the level of CCSD(T).
Therefore, we instead utilize the transfer-learning approach to obtain a computationally efficient methodology, which was first applied to the proton-tunneling dynamics of malonaldehyde.\cite{MM.tl:2022}
Using extensive sampling, a low-level NN-PES was constructed from MP2 energies, gradients and dipole moments.
The instanton prediction for the tunnelling splitting on this low-level PES was 96.3\wav, compared with 21\wav\ from the measurements.
However, by refining the PES based on only 25--50 CCSD(T) calculations at selected structures on or near the instanton path, the result dramatically improved to about 24\wav.
Adding the perturbative corrections brought the theoretical result to 22\wav, in excellent agreement with the experiment.\cite{MM.acc:2024} It was also found that reducing the low-level model to either DFT or even Hartree-Fock levels still allows to elevate to CCSD(T)/aug-cc-pVTZ quality through transfer learning to obtain accurate tunneling splittings.\cite{MM.tl:2023}

As a cutting-edge application tunneling splittings were determined for tropolone.\cite{MM.tl:2025}
which is chemically related to malonaldehyde but with 15 rather than 9 atoms. This almost doubles the dimensionality and
makes constructing a reactive PES significantly more complicated.
Brute-force construction of a NN-PES is unfeasible because each CCSD(T)/aug-cc-pVTZ/VTZ calculation (AVTZ for the 3 atoms directly involved in H-transfer and VTZ for all other atoms) now takes about 50 hours (using 4 processors, 6 GB memory, and 200 GB disk space) which prevents calculation of the required $\sim 10^4$ energies and gradients. For this reason, it was necessary to reduce the number of high-level calculations to about 25 carefully selected points.
Again, the combination of transfer learning with perturbatively corrected instanton theory
obtained excellent agreement with experiment,
beyond what had been achieved in previous theoretical attempts.\cite{Houston2020tropolone,tropolone} The calculated splittings $\Delta_{\rm RPI} = 1.07$ cm$^{-1}$ and  $\Delta_{\rm RPI+PC} = 0.94$ cm$^{-1}$ compare very favourably with the measurements $\Delta_{\rm Expt.} = 0.974$ cm$^{-1}$.\cite{tanaka:1999} Interestingly, the VPT2 vibrational spectrum from this TL-PES is also in excellent agreement with experiment.

As a final example, the TL-approach was applied to oxalate for which the IR-spectrum has been measured\cite{wolke:2015} but no tunneling splittings were measured so far. In other words, tunneling splittings are now {\it predicted}. The expermimentally determined IR-spectrum features an unusually broad band between 2000 and 3000 cm$^{-1}$ and was qualitatively reproduced in earlier MD studies.\cite{MM.oxa:2017} However, the TL-PES at the CCSD(T)/aug-cc-pvtz level of theory unambiguously assigned this particular spectral feature to the H-transfer motion.\cite{MM.oxalate:2025} In addition, the relative intensities and positions of the framework modes (below 2000 cm$^{-1}$) were in excellent agreement with the measured bands. Based on this validated ML-PES, the {\it predicted} H-transfer splitting is  $\Delta_{\rm RPI+PC} = 35$ cm$^{-1}$.

\section{Conclusions and Outlook}
In summary,
we have presented examples for how high-accuracy simulations of molecules and chemical reactions can be performed using a combination of machine learning with semiclassical approximations to quantum dynamics.
In future work, we plan to go beyond the approximations of instanton theory
using path-integral molecular dynamics.\cite{PIMDtunnel,malonaldehydePIMD}
In principle, this method allows for the exact calculation of tunnelling splittings (within the errors of the underlying PES).
Although it is significantly more expensive than the instanton approach,
the computational cost can be met due to the development of efficient ML-PESs.
In this way, we can have a direct test of the accuracy of the electronic-structure methods and the machine-learning procedures against experimental measurements from high-resolution spectroscopy.

In addition, the combined approach can be extended to 
both perturbatively-corrected instanton rate theory \cite{PCIRT}
and
nonadiabatic chemical reactions using golden-rule instanton theory.\cite{GRperspective}
In principle, machine learning can be used not just for potential energy surfaces, but also for nonadiabatic couplings.
There is, however, an extra complication that nonadiabatic couplings are double valued (just as $x^2=4$ has two solutions, $x=\pm2$) and change sign after winding around a conical intersection.
This complication can be avoided by evaluating the outer product of the nonadiabatic coupling vector with itself, resulting in a single-valued entity that can be learned using standard methods.
\cite{MLNACs,Dupuy2024MLNACs}

The rapid evolution of the interface between ML and
computational chemistry represents a powerful symbiosis rather than a
replacement of traditional theory. By learning and interpolating
potential energy surfaces, ML models allow researchers to bypass
computationally expensive calculations, granting access to complex
molecular systems that were previously out of reach. However, such
data-driven acceleration works best when coupled with rigorous
theoretical frameworks that include the underlying
physics.
In particular, we need high-accuracy electronic-structure theories to generate the training data and rigorous semiclassical theories to determine the dynamics on the resulting ML-PESs.
Ultimately, while ML dramatically reduces the computational
cost, the expertise of the theoretician remains essential to correctly
define physical problems, develop improved theories, and interpret the new phenomena revealed by
such hybrid approaches.

\section*{Acknowledgements}
JD and JOR acknowledge financial support from the Swiss National Science Foundation through project 207772 ‘Nonadiabatic effects in chemical reactions’. The work of MM is financially supported by the Swiss National Science Foundation through
grants $200020\_219779$, $200021\_215088$, and the University of
Basel (MM), which is gratefully acknowledged.
\section*{Author Contributions}

\bibliography{references,refs-mm}

\end{document}